# The Conception of the Atom in Greek and Indian Physics

Roopa Narayan

**Abstract.** This note contrasts the Greek and Indian conceptions of the atom. It is shown that these conceptions are quite different in spirit.

## Introduction

This note is a brief comparison of the subtler concepts of the atomic theory of Greek [1-3] and India [4-7] in order to show how they are very different in essence. The indivisibility of atom in the Greek conception was taken to be axiomatic, whereas in the Indian conception the atom was not material and, therefore, not subject to further division. The reason why this happened is because the two conceptions were situated in very different world-views [7].

## Greek Conceptions

Thales of Miletus (624 to 545 B.C.), generally considered the first Greek philosopher, believed everything has a common underlying material to which all things return after death, though they have the ability to appear variously – a kind of material monism and water is the fundamental element. Earth floats in water like a piece of wood, all things have gods in them, and magnets have souls because they are sources of motion.

Anaximander who is roughly from the same period defined a neuter infinity which is unbounded from which everything emerges in pairs like hot-cold, etc which finally merge back in to the same infinity. Anaximenes described air as the basic substratum from which all matter evolves and dissolves back in to it. Heraclites may have been the first one to introduce the paradox of a fundamental unity which becomes many during the vast process of things happening in the universe. Pythagoras believed to be his successor proclaimed that air, fire, water or earth were not the underlying unity of the cosmos, it is the geometrical ratios, numbers and certain symmetry.

Xenophanes too pursued the problem of a unity which becomes a multiple observed reality of the world. Parmenides known to be his student described this unity as spherical and stated that since nothing can either come from nothing or disappear in to nothing, everything must stay as is. He was followed by Empedocles who seems to have summarized the ideas of all his predecessors with a cyclical time based on four elements.

Zeno introduced his paradoxes, but all these people generally believed in a common material unity giving rise to different kinds of things because of opposite qualities like hot, cold, etc which are what we sense. All these theories had a greater focus on the cosmic unity with no true emphasis on atomic theory until Leucippus and Democritus.

Leucippus about whom little is known along with his student Democritus (460 BCE) founded the atomic theory in order to describe the universe in accordance with the senses of motion and changes. Only two different elements- the void and the solid atoms which are infinite in number are used to explain the entire universe. The atoms themselves have a basic nature of sweetness, coldness, etc and the amount of void in between atoms accounts for the lightness or heaviness of materials [6]. Yet, Leucippus is known to have said that everything in the world happens for a reason. He and his student Democritus proposed the atom – which is further indivisible only to overcome Zeno's paradox of any magnitude being divisible to infinite parts where in each part can be further divided leading to an infinite regression.

Democritus has mentioned two kinds of knowledge, a genuine one which comes with reasoning and the non-genuine one which is associated with senses. His atoms did not have properties like color, sweetness, etc but these were acquired as a result of the motion and arrangement of the atoms [2, page 66]. Yet these atoms with different shapes are not interchangeable [2, page 71], while in Indian physics there is only one kind of atom so to speak but it can have four basic kinds of motions which cannot co-exist in the same atom. In general this idea along with atomicity and unity sounds similar to Indian atomicity though the finer details and elaborateness of Indian atomic theory help see the difference between the two.

These early Greek atomists who along with other pre-Socratic philosophers like Empedocles and Anaxagoras proposed that there are multiple unchanging atoms – which means indivisible- which merely rearrange in order to form different materials. The differences in these material qualities like smell, color, etc were only due to interaction of matter with the senses. This mapping of matter to senses is found in $17^{th}$ – 18th century by Voltaire and D'Alembert [3, page 128]. These atoms are unchangeable, un-generated and indestructible moving in infinite void. Atoms and void are the two constituents of the universe which is created by god. This same idea of atoms is later accepted by Gassendi and Robert Boyle in $17^{th}$ century and the cause of motion of these atoms is attributed to God [3, page 127].

Plato's ideas about things are mentioned in his two statements where he states that each essence is the essence of exactly one form. Each form has (or is) exactly one essence. The physical word is a copy of this form and hence dependent on it and are deficient in comparison to these forms. But he is known to have disliked atomic theory and Democritus as well [2, page 67]. Aristotle too follows on the essence concept in relation to substance and does not propose any direct atomic theory though he wrote about previous atomists.

Epicurus (341–270 B.C.) closely followed Democritus with a physical theory of atoms but his theory was devoid of God. All the atoms are of finite number of different shapes and sizes but too small to be seen all moving in the same speed. Everything can be deduced from observation and hence he too concludes things can neither come in to existence from nothing nor get reduced in to nothing. Since objects and empty space are the two kinds of realities observed, the same is true at atomic level as well and solid objects being full offer resistance while empty space cannot do so. The finite number of objects seen lead to a finite kind of atoms but infinite in number.

## Indian atomic theory

Kanada is the founder of the Indian School of physics which is called Vaisheshika. He uses logic as a tool to deduce that all matter must be made of an indivisible entity which he calls *anu*-the atom (for details, see [7]).

In Indian physics matter is reduced to an abstract *anu* which emerges as four basic kinds of matter owing to four basic distinguished kinds of motion of this atom/*anu*. This atom is that in which the minima of magnitude rests in terms of volume, mass, etc and no kind of measure can be associated with it. Space is said to be "opposite" of this atom since in space rest the maxima of magnitude and irrespective of the method employed, both these are beyond perception.

The beginning of creation process is characterized by motion which is acquired by the atom along with which it acquires certain inherent properties and that is when time also begins. Kanada uses his categorization to reduce all matter, space and time to certain functions of 'motion'. In the absence of motion even time collapses to zero. The observer represented by the mind is also a function of motion.

The entire universe is only matter and the observing mind that are capable of motion. Indian Physics is an observer centered system with space-time as the fundamental matrix through which the entire universe is observed by the observer. It is set in the framework of a cyclic cosmological model dealing with endless creations and dissolutions. Time is said to collapse in the rest period between the cosmic creation and dissolution, and that must be true if time is a function of 'state of motion' of the cosmos which comes to a rest in this period between creations and dissolutions.

Matter is conserved in the atomic state. Space too is eternal and continues to stay as is. The *anu* of Indian physics is an atom to the extent that it is further indivisible, and is uncaused or indestructible. But the difference lies in the fact that Kanada does not make an attempt to describe visible matter alone, but instead proposes a complete system of space, time, matter, etc to describe the entire cosmos which begins with the visible matter and extends to categories and potentials that are not matter-like. His atom does not exist in the real-time. This atom in its fundamental form possesses no motion and motion by definition is visible.

## Conclusion

In the beginning of western science the observed world was a creation of an external God which became an objective reality to be studied [3, p 84]. In Indian physics the entire universe is categorized from the point of view of the observer. The relation between whole and parts has a different approach. The matter alone is divisible in to atoms and the maximum number of parts is reaches only at during the cosmic dissolution. Atom is the minima of magnitude whole space is the maxima of magnitude.

Heisenberg in his "The Physicist's Conception of Nature" summarizes the context of western science in the following words [3]:

> *This change in the scientists' attitude to nature is perhaps best understood if we consider that, to Christian thought of the time, God seemed to be in a heaven so high above the earth, that it became significant to look at the earth without reference to God.*

This science was not only independent of religion or God but also independent of man. Hence scientists tried to arrive at 'laws' through mathematical relationships which remained the same in the entire cosmos which helped harness nature's forces for the human purpose.

It is this central idea of the western science which is in contrast with the Indian physics, which from the beginning was based on the idea of the cyclic cosmos, which is alive in the sense that there is a constant interaction between the observer and the cosmos.